\documentclass{article}
\usepackage[utf8]{inputenc}
\usepackage{graphicx}
\usepackage[square,comma]{natbib}
\usepackage{amsfonts}
\usepackage{appendix}
\usepackage{authblk}
\usepackage{bbm}
\usepackage{url}
\usepackage{amsmath}
\usepackage{mathrsfs}
\usepackage{mathtools}
\usepackage{gensymb}
\usepackage{amssymb}
\usepackage{upgreek}
\usepackage{yhmath}
\usepackage{array}
\usepackage{xcolor}
\usepackage{ulem}
\usepackage{MnSymbol}

\title{Brief Communication: Dimensionality Reduction in Total Dynamic Mode Decomposition Using A Simple Geometric Method}

\author[1]{Christopher J. Keylock}

\affil[1]{School of Architecture, Building and Civil Engineering, Loughborough University, Leicestershire, LE11 3TU, UK. E-mail: c.j.keylock@lboro.ac.uk}

\begin{document}

\maketitle

\begin{abstract}
Dynamic mode decomposition (DMD) and its variants have emerged as popular methods for the post-processing of fluid dynamics' simulations in order to visualize dominant coherent structures and to reduce the practical degrees of freedom to a restricted set of ``modes''. In this brief communication we provide a geometric method for choosing the number of modes for the Total DMD technique and test its efficacy using a synthetic example (to examine the effect of noise) and a cylinder wake case.
\end{abstract}

\section{Dynamic Mode Decomposition and its Variants}

Since its original formulation \cite{schmid10}, dynamic mode decomposition (DMD) has emerged as a popular tool for processing ``snapshots'' in not only fluid mechanics, but a range of different disciplines \cite[e.g.][]{higham17,erichson19,ikeda22}. Theoretically oriented work has forged connections between the Koopman operator \cite{koopman31} and DMD \cite{rowley09,mezic12}, 
leading to extended DMD \cite{williams15}, which approximates the Koopman modes, eigenvalues and eigenfunctions, in contrast to standard DMD, which does not approximate all of the eigenfunctions (because it lacks a quadratic term - see also higher order DMD approaches \cite{vega18}). There have also been a number of other developments of the original method including optimal DMD \cite{wynn13} and sparsity promoting DMD \cite{schmid14}. The innovation that is central to this paper it total DMD \cite{hemati17}, which adds a pre-processing step to ensure that the error in decomposition is distributed across both companion matrices (a more robust approach than projecting all of the error into one given that these matrices only differ from each other by one column). 

To summarise the nature of dynamic mode decomposition, we assume we have a set of data vectors $\{x_{0},x_{1},x_{m}\}$ representing the physics of the process. Typically, these will represent states of the system measured at $n$ different locations at different times, and will be in time order with a constant separation, $\Delta t$, although a sequential ordering is not necessary for all variants of DMD. One then seeks a linear model, $x_{k} = \mathbf{A} x_{k-1}$. In order to estimate this model we first form the two companion matrices, $\mathbf{X} = \{x_{0},x_{1},x_{m-1}\}$, $\mathbf{Y} = \{x_{1},x_{2},x_{m}\}$. Thus the estimation problem is $\mathbf{Y} = \mathbf{A}\mathbf{X}$ and the original approach to this was to undertake a singular value decomposition (SVD) of $\mathbf{X}$ \cite{schmid10}:
\begin{equation}
\label{eq.SVD}
    \mathbf{X} = \mathbf{U}\boldsymbol{\Sigma}\mathbf{V}^{*},
\end{equation}
where $\mathbf{U}$ is $n \times r$ and contains the left singular vectors, $\boldsymbol{\Sigma}$ is $r \times r$ and contains the singular values on the diagonal in descending order, and $\mathbf{V}$ is $r \times n$ and contains the right singular vectors, where $r \le m$ is the rank of $\mathbf{X}$. Dimensional reduction is then used to restrict attention to the first $N$ of these $r$ modes \cite{lumley70}. Our linear operator, $\mathbf{A}$ is then estimated using the restricted SVD and the Moore-Penrose pseudo-inverse:
\begin{equation}
\label{eq.Atilde}
    \tilde{\mathbf{A}} = \mathbf{U}^{*}\mathbf{Y}\mathbf{V}\boldsymbol{\Sigma}^{-1}.
\end{equation}
The eigenvectors for $\tilde{\mathbf{A}}$ are then given by
\begin{equation}
   \tilde{\mathbf{A}}e = \lambda e,
\end{equation}
and the original DMD then gave the \textit{ projected} DMD modes as the values for $\mathbf{D}$ corresponding to non-zero eigenvalues, $\lambda$:
\begin{equation}
\label{eq.D}
   \mathbf{D} = \mathbf{U} e, 
\end{equation}
where the eigenvectors, $e$ are usually scaled to a unit norm. The projected DMD modes are an optimal representation of $\mathbf{A}$ in a basis formed from the Proper Orthogonal Decomposition modes of $\mathbf{X}$ \cite{schmid14}. A refinement to the formulation of the modes themselves, but built on the same dimensionality reduction step is \textit{exact} DMD \cite{Tu14} where (\ref{eq.D}) is replaced with the following expression for the \textit{i}th exact DMD mode:
\begin{equation}
\label{eq.exactD}
D_{i}^{(ex)} = \frac{1}{\lambda_{i}} \mathbf{Y} \mathbf{V} \boldsymbol{\Sigma}^{-1} e_{i}. 
\end{equation}

Irrespective of if one makes use of (\ref{eq.D}) or (\ref{eq.exactD}) to recover the modes, the total DMD approach for distributing the error across both companion matrices is advantageous. I.e. rather than a least-squares estimation of $\tilde{\mathbf{A}}$, total least squares estimation is applied \cite{hemati17}. This approach is implemented by defining the augmented matrix
\begin{equation}
    \mathbf{Z} = \left[ \begin{array}{c}\mathbf{X}\\ \mathbf{Y}\end{array} \right],
\end{equation}
applying an SVD to $\mathbf{Z}$ to give $\mathbf{P}\mathbf{L}\mathbf{R}^{*} = \mathbf{Z}$, applying dimensionality reduction to $\mathbf{R}$ to restrict it to the first $N$ right singular vectors and then using $\mathbf{R}^{(N)}$ to project into a common basis for the new companion matrices: 
\begin{align}
\label{eq.TDMD}
\nonumber
\hat{\mathbf{X}} &= \mathbf{X} \mathbf{R}^{(N)}\\ \hat{\mathbf{Y}} &= \mathbf{Y} \mathbf{R}^{(N)}.
\end{align}

\section{The Dimensionality Reduction Criterion}
The aim of this communication is to examine the dimensionality reduction step in total DMD, i.e. the move from $\mathbf{R}$ to $\mathbf{R}^{(N)}$. Our idea is based on the fact that the  multiplication by $\mathbf{R}^{(N)}$ in (\ref{eq.TDMD}) imposes an ordering based on the structure of the singular values in $\mathbf{L}$. Hence, one expects the left-most columns of $\hat{\mathbf{X}}$ and $\hat{\mathbf{Y}}$ to be highly correlated, and for this correlation to decline as the singular values decline in magnitude and become more affected by intrinsic variability or extrinsic noise. Formulating this correlation as the angle between the $i \in \{1,\ldots,N^{/}\}$ vectors in $\hat{\mathbf{X}}$ and $\hat{\mathbf{Y}}$ \cite{golub73} we have
\begin{equation}
\label{eq.theta}
\theta_{i} = \cos(\overrightarrow{\hat{X}_{i},\hat{Y}_{i}}).
\end{equation}
Our criterion is then essentially to choose $N$ as the cardinality of the set for which $\theta_{i} > sqrt{2}/2$, i.e. the departure between $\hat{X}_{i}$ and $\hat{Y}_{i}$ is less than the physically meaningful value of $45\degree$. The necessary refinement to this statement is a consequence of noise, which we absorb into the threshold definition based on the departure of $\theta_{1}$ from 1. I.e. the threshold is given by $T_{1} = \cos(\text{acos}(\theta_{1}) + \pi/4)$.
However, as the amplitude of the noise increases and $\theta_{1}$ declines towards $\sqrt{2}/2$ then $T_{1}$ is inappropriate as it gives a values lower than may arise by chance. To deal with this we can bootstrap a lower limit for the threshold. That is, given two matrices, $\mathbf{B}_{X}$ and $\mathbf{B}_{Y}$ that are the same size as $\mathbf{X}$ and $\mathbf{Y}$ but consist entirely of independently and identically distributed normally distributed random values, we form $\tilde{\mathbf{B}}_{X} = \mathbf{B}_{X} \mathbf{R}^{(N)}$ and $\tilde{\mathbf{B}}_{Y} = \mathbf{B}_{Y} \mathbf{R}^{(N)}$ and then find the cosine of the angle between all $N^{/}$ vectors in a similar vein to (\ref{eq.theta}). The lowest limit for the threshold is then the maximum value over these $\tilde{\theta}$. In practice we can repeat this $1/\alpha$ times and find the maximum of these maxima such that the $T_{0}$ is a one-sided confidence limit on the maximum value at the $alpha$ significance level. (Here we assume the classical choice of $\alpha = 0.05$). Thus, we then have that
\begin{equation}
\label{eq.N}
    N = \left\{\begin{array}{l l}\mbox{card}\left[\theta_{i} > \max\left(T_{0}, \, T_{1}\right)\right] & \text{if}\,\, \theta_{1} > \sqrt{2}/2\\
    1 & \text{if}\,\, \theta_{1} \le \sqrt{2}/2\end{array}\right..
\end{equation}

\section{Testing the Criterion}

\begin{figure}
\noindent\includegraphics[width=33pc]{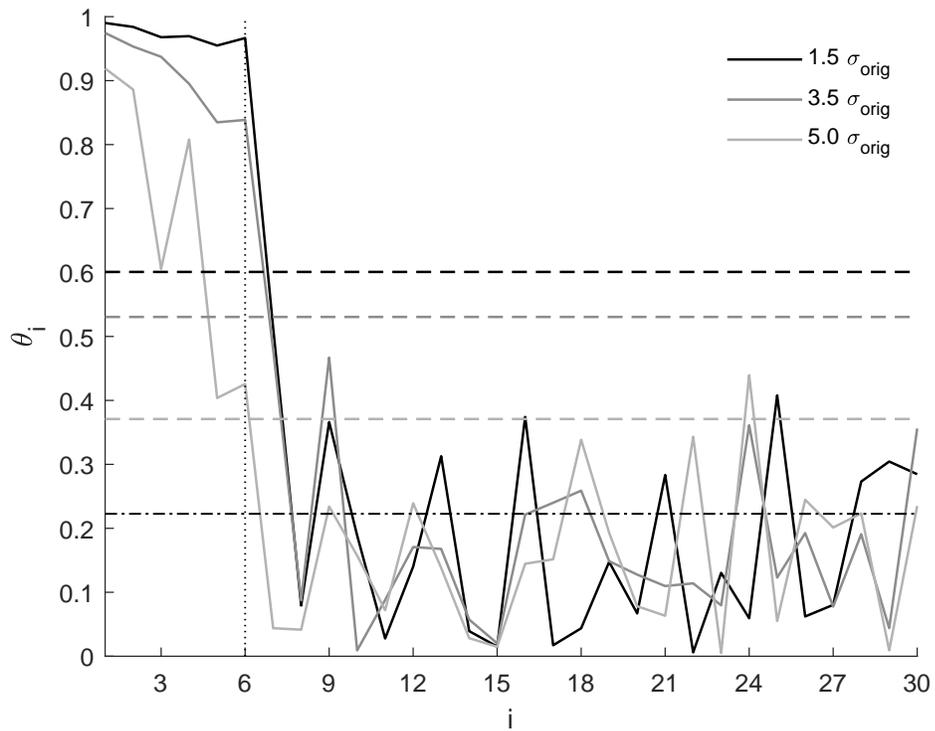}
\caption{A plot of $\theta_{i}$ versus $i$ for the ideal case considered in \cite{hemati17} but with three noise levels shown. The vertical dotted line at $i = 6$ shows the number of modes that should be recovered. The horizontal dot-dashed line indicates the bootstrapped lower limit for the threshold based on random matrices. The horizontal dashed lines are the thresholds corresponding to the three different cases, decreasing in value as $\theta_{1}$ decreases.}
\label{fig.rowley}
\end{figure}

To test our criterion for the effect of noise we use the synthetic ``Rowley System'' \cite{hemati17}, which consists of three pairs of dynamics modes, two of which are purely oscillatory while the third has some damping: $\lambda_{1} = \exp [(\pm 2\pi i)\delta t]$; $\lambda_{2} = \exp [(\pm 5\pi i)\delta t]$; $\lambda_{3} = \exp [(-0.3 \pm 11\pi i)\delta t]$, where $\delta t = 0.01$s. A linear mapping is then used to expand from $r = 6$ to $n = 250$ (i.e. the snapshot contains 250 ``pixels'') and then a time-series of $m = 100$ is adopted. We define the amount of noise in multiples of the standard deviation of the values in the initial snapshot matrix, $\sigma_{orig}$ (Hemati et al. chose a value for the noise of $\mathcal{N} \sim 1.5 \sigma_{orig}$ \cite{hemati17}). This is the case shown by the black line in Fig. \ref{fig.rowley} and the correct number (six) of dynamic modes is clearly extracted for each of these cases. For $\mathcal{N} \ge 7.5 \sigma_{orig}$, $\theta_{1} < \sqrt{2}/2$ and noise has corrupted the relation between $\hat{\mathbf{X}}$ and $\hat{\mathbf{Y}}$. For $5 \,\sigma_{orig} < \mathcal{N} < 7.5 \,\sigma_{orig}$ the number of modes found declines from 6 to 1 as the noise amplitude increases.  

\begin{figure}
\noindent\includegraphics[width=33pc]{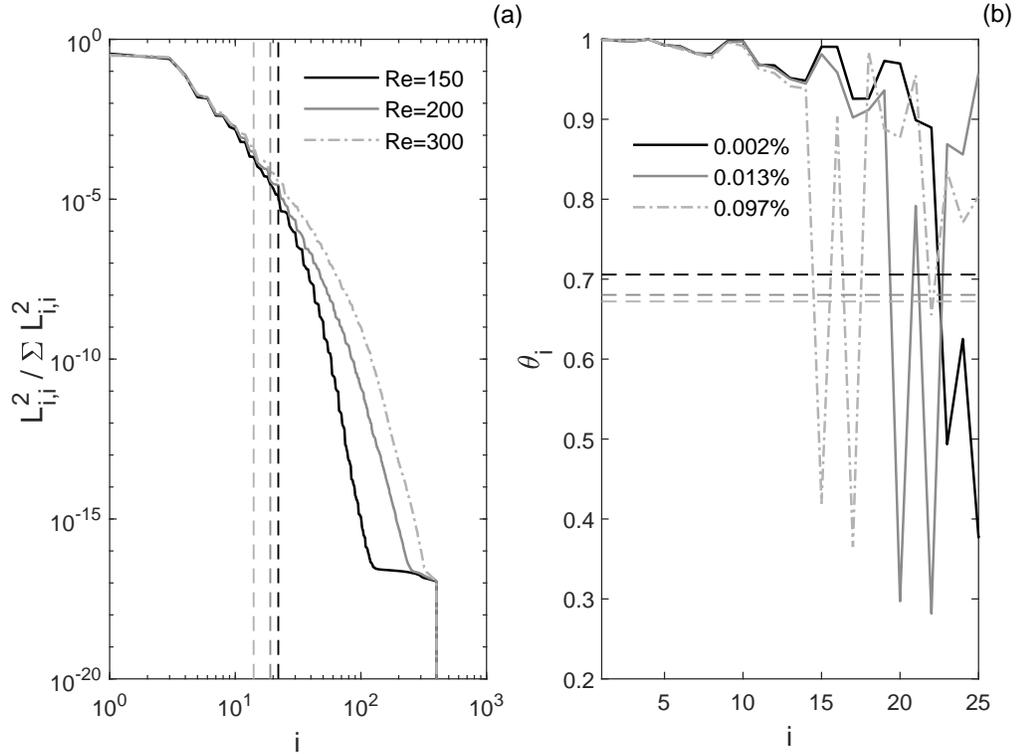}
\caption{The square of the singular values as a proportion of the total for the cylinder wake data at three Reynolds numbers are shown in panel (a). Panel (b) shows $\theta_{i}$ versus $i$ for $i \le 25$.}
\label{fig.cyl}
\end{figure}

We then examine three cylinder wake cases studied by Cai and co-workers\cite{cai19}. In each case we study the fluctuating flow field (the average snapshot is subtracted from each snapshot before transformation into a vector). Panel (a) indicates how the energy of the singular values decays with $i$, with the values for $N$ superimposed for each Reynolds number as vertical lines. Panel (b) shows $\theta_{i}$ against $i$ with the values for $T_{1}$ as horizontal lines. The percentages quoted are the residual energy in the singular values greater than $N$. I.e. for $\text{Re} = 150$, $99.998\%$ of the total energy is captured in the first $N = 22$ singular values. 

\begin{figure}
\noindent\includegraphics[width=33pc]{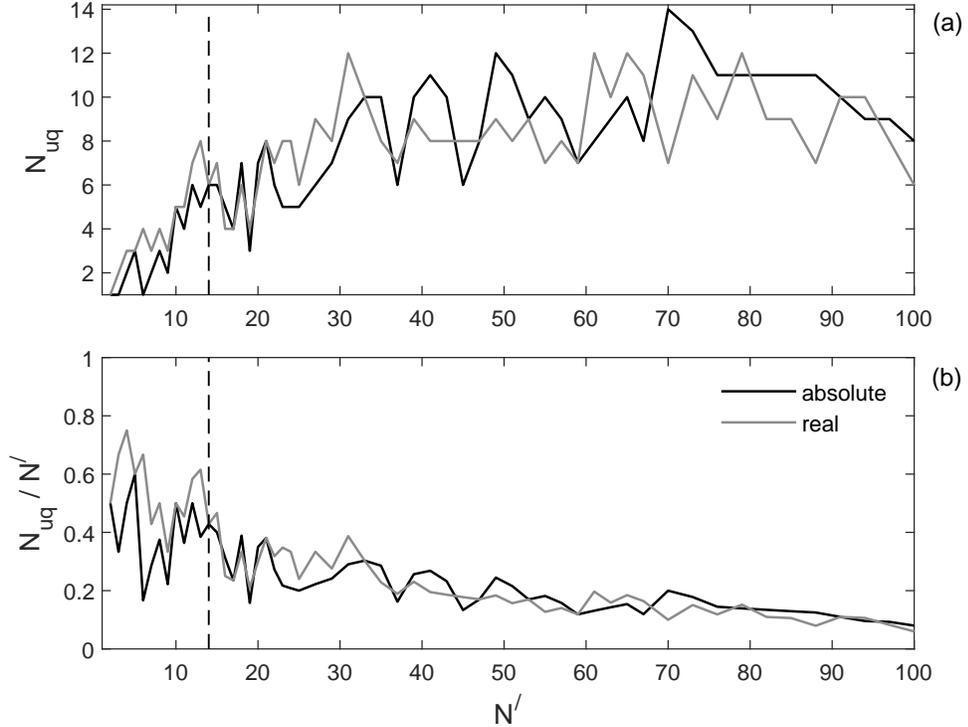}
\caption{The number of exact DMD modes uniquely contributing to the maximal correlations between a time snapshot and each of the $N^{/}$ exact modes for $2 \le N^{/} \le 100$. Panel (b) normalises the raw values by $N$. Results are shown for both the real part and the absolute value for each DMD mode. The vertical dashed line is the value for $N$.}
\label{fig.Nunique}
\end{figure}

Given that the highest Reynolds number is the case with the greatest residual variance, we focus on this case, and vary the number of exact DMD modes extracted from $2 \le N^{/} \le 100$. We then correlate each mode back against the snapshots in the original time-series and record the maximum absolute correlation for each frame and the mode responsible. Defining $N_{uq}$ to be the unique number of modes contributing to these maximal correlations, in Fig. \ref{fig.Nunique} we show the values for $N_{uq}$ as a function of $N^{/}$. We include results for correlations based on both the absolute value and real part of each mode, with results normalised by $N^{/}$ in panel (b). Both sets of results indicate a value for $N_{uq} \sim 6$ at our choice of $N$ (vertical, dashed line). Panel (a) shows that $N_{uq}$ is initially limited by the number of modes available. For $N^{/} > N$, the value for $N_{uq}$ increases somewhat indicating some potential value in using more modes than $N$. However, for $N^{/} \gtrsim 20$ the efficiency of including this number of modes has clearly decreased relative to $N = 14$.

\section{Conclusion}
The total DMD framework \cite{hemati17} is an important development in the evolution of dynamic mode decompositions, the use of which now extends well beyond fluid mechanics. Given that the primary value in such decomposition methods is in their reduction of dimensionality, our proposed criterion, which makes use of the particular singular value structure of total DMD (\ref{eq.N}) but contains minimal additional assumptions, extracts the correct number of modes for an idealised problem even with significant noise corruption (Fig. \ref{fig.rowley}). Testing using a cylinder wake test dataset \cite{cai19} restricts the number of modes from potentially 500 down to $14 \le N \le 22$ depending on Reynolds number. When correlating the modes back to the original snapshots, our criterion would appear to be efficient in terms of extracting the correct number of modes to correlate maximally with the information in the original data. We hope this criterion can be of generic use in fluid mechanics' post-processing.


\section{Compliance with ethical standards}

Conflict of Interest: The author declares that he has no conflict of interest.

Funding: There is no funding source.

Ethical approval: This article does not contain any studies with human participants or animals.

Availability of data and materials: The cylinder wake data are taken from \cite{cai19} and are available at:

https://github.com/shengzesnail/PIV\_dataset/tree/master/PIV-genImages.

\bibliographystyle{spmpsci_unsort}       
\bibliography{JFMk}   

\end{document}